 \documentclass[final,3p,times]{elsarticle}

\usepackage{graphicx}

\usepackage{amssymb}
\usepackage{amsmath}

\journal{Proceedings of the Combustion Institute}

\begin{document}
\begin{frontmatter}



\title{Dynamics of shock induced ignition in Fickett's model with chain-branching kinetics: influence of $\chi$}


\author{J. Tang \& M. I. Radulescu}

\address{Department of Mechanical Engineering, University of Ottawa, Ottawa Canada K1N6N5}

\begin{abstract}
The problem of shock induced ignition by a piston is addressed in the framework of Fickett's model for reactive compressible flows, i.e., the reactive form of Burgers' equation.  An induction-reaction two-step chain-branching model is used to study the coupling between the energy release and the compressible hydrodynamics occurring during the shock ignition transient leading to a detonation.  Owing to the model's simplicity, the ignition and acceleration mechanism is explained using the two families of characteristics admitted by the model.  The energy release along the particle paths provides the amplification of forward-travelling pressure waves. These waves pre-compress the medium in the induction layer ahead of the reaction zone, therefore changing the induction delays of successive particles. The variation of the induction delay provides the modulation of the amplification of the forward travelling pressure waves by controlling the residence time of the pressure waves in the reaction zone.  A closed form analytical solution is obtained by the method of characteristics and high activation energy asymptotics.  The acceleration of the reaction zone was found to be proportional to the product of the activation energy, the ratio of the induction to reaction time and the heat release.  This finding provides a theoretical justification for the previous use of this non-dimensional number to characterize the ignition regimes observed experimentally in detonations and shock induced ignition phenomena. Numerical simulations are presented and analysed.  Both subsonic and supersonic internal flame propagation regimes are observed, consistent with experiment and previous reactive Euler models.   
\end{abstract}


\end{frontmatter}
\section{Introduction}
\addvspace{10pt}
The present study focuses on the problem of shock induced ignition by a piston suddenly accelerated into a reactive medium.   This problem has attracted significant interest in recent years due to its universality in capturing the canonical processes occurring in reactive media in the presence of gas-dynamic disturbances, i.e. pressure waves \cite{Clarke&Cant1985, Blythe&Crighton1989,  Bauwens2000,  Blytheetal2009, Clarkeetal1990, Dold&Kapila1991, Jackson&Kapila1985, Sharpe&Short2007, Sharpe&Short2004, Sharpe2002}.  The problem is also equivalent to shock induced ignition by a reflected shock \cite{Strehlow1968}.  

Of particular interest to us is to determine the parameter that controls the different one-dimensional ignition regimes that can be realized experimentally \cite{Strehlow1968}.  For example, Meyer and Oppenheim first suggested that the sensitivity of the induction time to temperature fluctuations, with respect to the duration of the exothermic power pulse controls the ignition process \cite{Meyer&Oppenheim1971}.  Mathematically, this can be written as the product of the reduced activation energy controlling the induction reactions and the ratio of induction to reaction times \cite{Radulescu2003}, namely,  
\begin{align}
\chi=\frac {t_{ind}}{t_{reac}} \frac{E_a}{RT_s} \label{eq:chai}
\end{align}
The same parameter was also identified in the stability analyses of self-propagating detonation waves \cite{Short&Sharpe2003, Ngetal2005, Leungetal2010} and has also been used for correlating the ignition regimes and stability of cellular detonations \cite{Radulescu2003}.  Typically, a value of $\chi \approx 10$ corresponds to the onset of strong pressure waves driven during ignition behind the shocks and the loss of stability of detonation waves.  However, the physical mechanism responsible for changes of the shock-ignition regime when this parameter is varied remain unclear. 

To address the relative importance of the induction zone sensitivity and the duration of the exothermic stage, the present study assumes the medium's decomposition is via a two step chain branching model, with distinct induction and reaction stages. This is typical of most reactive gases and can be shown to rationally derive from a generic three step chain branching model \cite{Blytheetal2006, Sharpe&Maflahi2006}. The investigation thus builds upon the studies of Dold and Kapila \cite{Dold&Kapila1991}, Sharpe \cite{Sharpe2002} and Blythe et al. \cite{Blytheetal2009}, who have addressed this problem in the past using the reactive Euler equations coupled with two and three-step chain-branching models.   

Although significant insight has been gained in these past studies, the present work focuses on the much simpler Fickett model for compressible reactive media \cite{Fickett1979, Fickett1985}.  This model is an extension of Burgers' equation, a prototypical model for inert compressible hydrodynamics, to a reactive medium.  Due to its simplicity, this model can serve to study the basic mechanism without the complications of the highly non-linear reactive Euler equations.  For example, this simple model has allowed us in the past to clarify the instability mechanism of pulsating detonations and has reproduced the universal period doubling route to chaos in travelling detonations \cite{Radulescu&Tang2011}.  Using this model, we attempt to determine in the simplest form possible how the $\chi$ parameter controls the dynamics of the ignition process following shock compression.
\section{Fickett's reaction model}
\addvspace{10pt}
Fickett's analogue model to one-dimensional inviscid reactive compressible flow is given by \cite{Fickett1985, Radulescu&Tang2011}:
\begin{align}
\partial_t \rho + \partial_x p=0 \label{eq:Fickett}
\end{align}
The model is formulated in Lagrangian coordinates, $x$ denotes a material coordinate and $t$ represents time \cite{Fickett1979, Fickett1985}.  The variable $\rho$ has the meaning of density in the model. The term $p$ in \eqref{eq:Fickett} has the meaning of pressure, see \cite{Fickett1985}. For simplicity, we choose the generic equation of state proposed by Fickett, that is
\begin{align}
p=\frac{1}{2} \left(\rho^2 + \lambda_r Q \right) \label{eq:pressure}
\end{align}
where $\lambda_r$ is a reaction progress variable ranging from 0 (unreacted) to 1 (reacted).  The model has the property that pressure increases with increasing density, or release of energy at constant density.  Note that one recovers the inviscid Burgers' equation by setting $Q=0$.  A similar simplified reactive flow model was also proposed by Majda \cite{Majda1981}, with slight differences in formulation.

In the present work, we adopt the two step reaction model proposed by Radulescu and Tang \cite{Radulescu&Tang2011} in order to capture the main features of chain-branching reactions \cite{Fickett&Davis1979}.  Following the leading shock, we assume the existence of a thermally neutral induction zone, whose duration depends on the medium's local compression.  Following the induction stage, we assume an exothermic reaction that proceeds at a state-independent constant rate.  The resulting generic induction-reaction model we propose is thus:

\begin{align}
\partial_t \lambda_i = -H(\lambda_i) e^{\left( \frac{\rho - 1}{\epsilon} \right)} \label{eq:rate1}\\
\partial_t \lambda_r \equiv r =  K \left(1-H(\lambda_i)\right) \left(1-\lambda_r \right)^{\nu}  \label{eq:rate2}
\end{align}

The variables have been non-dimensionalized by the characteristic state behind the initial shock.  Time is thus non-dimensionalized by the induction delay behind the initial shock.  The parameter $\epsilon$ is the inverse activation density of the induction reactions. The parameter $K$, through our non-dimensionalization, is the ratio of the induction and reaction times.  The Heaviside function $H(\cdot)$ controls the timing of the onset of the second exothermic reaction, which starts when the induction variable $\lambda_i$ reaches 0.  Ahead of the shock, $\lambda_i=1$ and  $\lambda_r=0$.  We are also assuming that the induction reaction is activated by the passage of the inert leading shock.  The system to be solved is thus \eqref{eq:Fickett}, \eqref{eq:rate1} and \eqref{eq:rate2} with the equation of state given by \eqref{eq:pressure}.

For the piston initiation problem of interest, the piston path corresponds to $x=0$ (see Fig. \ref{fig:K=0.2}).  In order to model a steady piston, we set $\rho(x=0,t)=1$.  Without loss of generality, we assume the strong shock limit and take the quiescent gas ahead of the shock as $\rho(x,t=0)=0$, $\lambda_i(x, t=0)=1$ and $\lambda_r(x, t=0)=0$.  For a non-reactive medium, Burgers' equation with the prescribed initial and boundary conditions result in a constant strength shock, propagating with speed $1/2$, followed by a constant state $\rho=1$ \cite{Whitham1974}.  When the medium is exothermic, the energy which is deposited in the medium alters the medium's mechanical response; we wish to analyse this mechanical response below.
\section{Amplification mechanism}
\addvspace{10pt}
We first solved the above system numerically by a finite volume technique described by Radulescu \& Tang \cite{Radulescu&Tang2011}. In the present study, a grid resolution of 1600 grid points per unit length was used.  We also set $Q=1$.
 
\begin{figure*}
\begin{center}
\includegraphics[width=110mm]{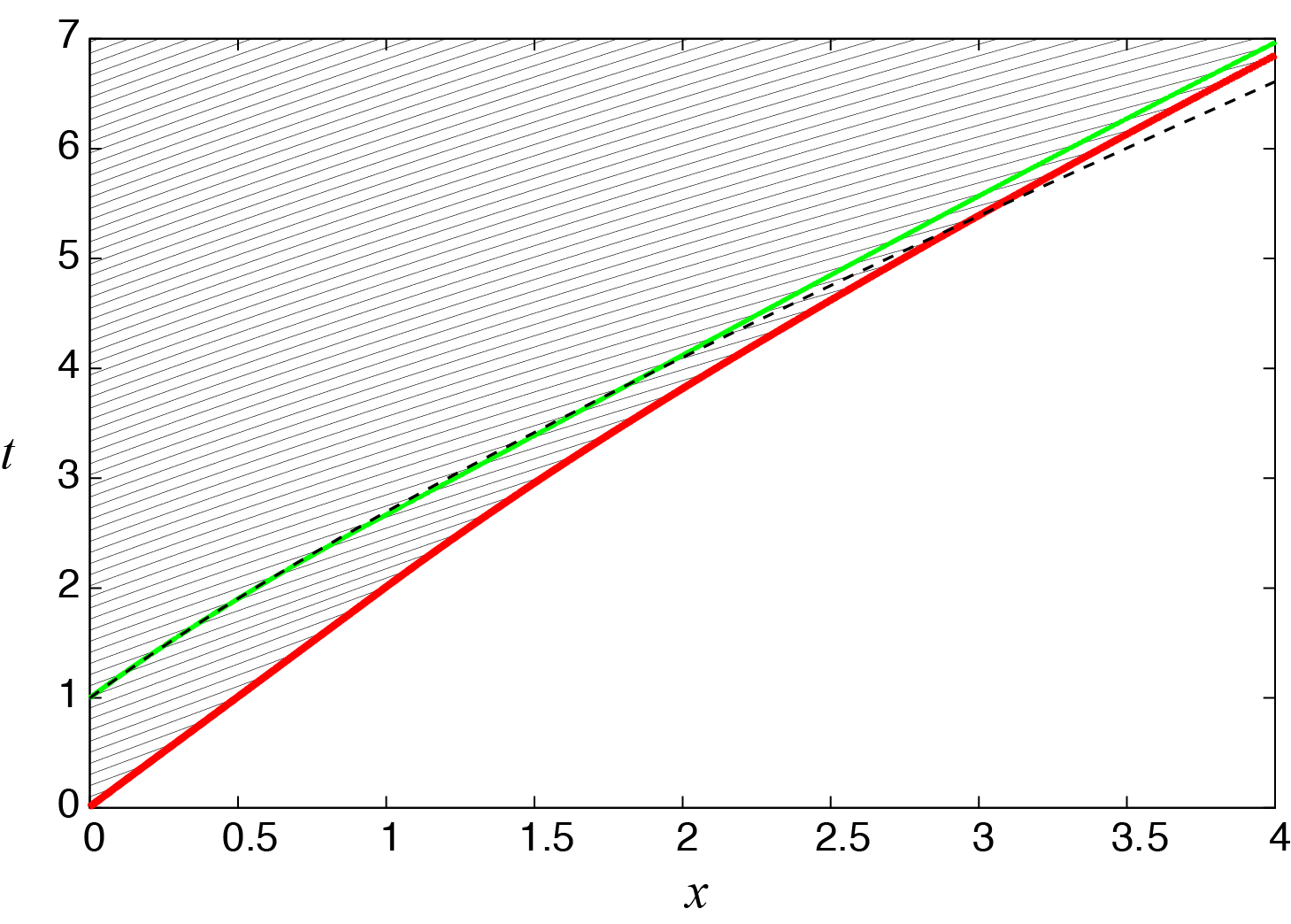}
\caption{Characteristic $t-x$ plot of shock initiation for $K$=0.2, $\epsilon$=0.2; the red line represents the main shock front $t_s(x)$, the green line represents the fire trajectory $t^*(x)$; the solid lines are $C^+$ characteristics; the broken line is the analytical prediction for the fire trajectory given by \eqref{eq:tstarfinal} for $t=O(1)$.} 
\label{fig:K=0.2}
\end{center}
\end{figure*}

Figure \ref{fig:K=0.2} shows an example of the solution obtained numerically for $K=0.2$ and $\epsilon=0.2$.  The variation of $\rho$ is shown in the accompanying \textit{Video 1}, but can be deduced directly, as it is given by the slope of the pressure waves (see below).  The trajectories of the leading shock (the red line in the figure), the onset of energy release, which we will henceforth call \textit{fire} (green line), and the internal pressure waves (black lines) are shown in the \textit{t-x} diagram.  The pressure wave emitted at the onset of reactions at $(x=0, t=1)$ marks the end of the inert solution.  The subsequent solution consists of an accelerating shock and fire, which eventually form a detonation wave.  

The physical mechanism responsible for the amplification shown in Fig. \ref{fig:K=0.2} is best explained by first transforming the governing equations in characteristic form and integrating \eqref{eq:rate1} from the time a particle crosses the shock at $t=t_s(x)$ to the time at which the exothermic reactions begin at the fire $t=t^*(x)$.  We obtain:
\begin{align}
\frac{d p}{d t}=\frac{1}{2}rQ \; \mathrm{\quad along \quad}\;  \frac{dx}{dt}=\rho \label{eq:charplus}\\
1=\int_{t_s}^{t^*}exp\left( \frac{\rho-1}{\epsilon} \right) \label{eq:indintegral}\\
\frac{d \lambda_r}{d t}=r \; \mathrm{\quad along \quad} \; \frac{dx}{dt}=0\label{eq:char0}
\end{align}

By inspection of \eqref{eq:charplus}, we deduce that the pressure waves propagating forward with acoustic speed $dx/dt=\rho$ amplify whenever they travel through a reactive field.  This is the case of the pressure waves in Fig. \ref{fig:K=0.2} originating at the piston once the reactions have begun at $t=1$.  The rate of change of the pressure wave's amplitude varies with the heat addition $Q$ at the rate $r$.  Indeed, Figure \ref{fig:K=0.2} shows these characteristics accelerating forward.  For a constant reaction rate, the pressure waves amplification is proportional to their residence time in the reaction zone.  Once these waves exit the reaction zone, they propagate into the induction zone, compressing the unreacted material and strengthening the leading shock.  

The other key element in the amplification stage is the dependence of the induction delay on this pre-compression, via \eqref{eq:indintegral}.  For low $\epsilon$ (i.e., high activation energy), the induction delay can be significantly reduced.  The tendency towards lowering the ignition delay permits the pressure waves to remain in the reaction zone for longer times, as can be seen from Fig. \eqref{fig:K=0.2}.  This provides a more coherent amplification of these pressure waves, which in turn enhances the compression in the induction zone and that of the shock.  The feedback mechanism is thus between the pressure waves propagating forward and the modification of the onset of reaction, which modulates the amplification amount.  This mechanism was also found central in controlling the detonation wave stability by Radulescu and Tang \cite{Radulescu&Tang2011}.  

\begin{figure}
\begin{center}
 \includegraphics[width=110mm]{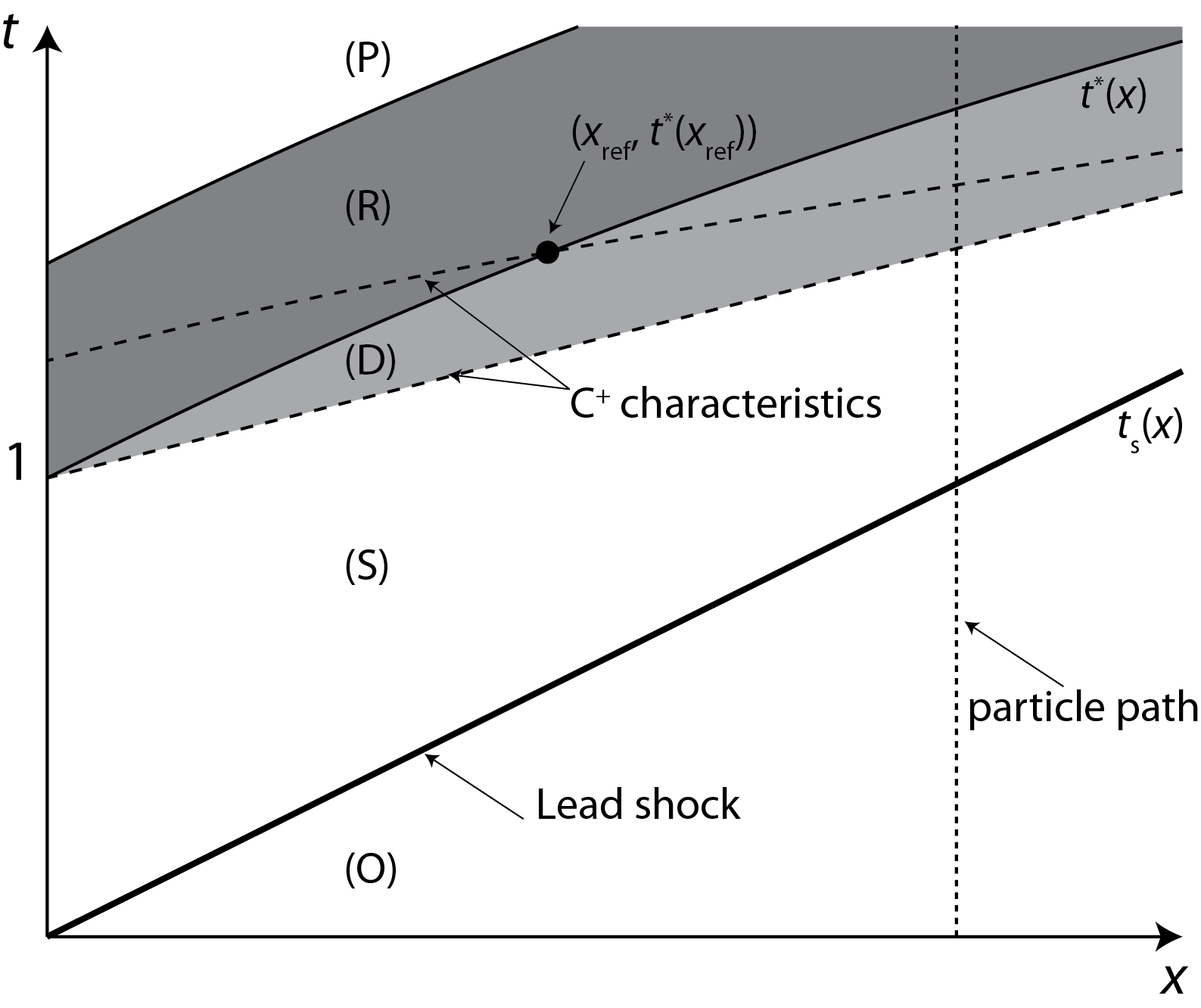}
\caption{Space-time diagram illustrating the run-away process at early times. A $C^+$ pressure wave originating at the piston $(x=0)$ amplifies while travelling through the reaction zone (R).  Its emergence into the induction zone modifies the post shock state from (S) to a (D).  This shortens the ignition delay ($t^*-t_s$).  The shorter ignition delay in turn provides a more coherent amplification of $C^+$ pressure waves in the reaction zone. Zones O and P denote the initial and product states, respectively.} 
\label{fig:fig1}
\end{center}
\end{figure}
\section{Asymptotic analysis}
\addvspace{10pt}
The early transient dynamics illustrated in Fig. \ref{fig:K=0.2} and discussed above have been summarized in Figure \ref{fig:fig1}.  The arguments presented can now be posed mathematically in an asymptotic analysis, which assumes that the inverse activation energy is small (i.e., $\epsilon<<1$).  The framework is similar to that used by Sharpe \cite{Sharpe2002} for the reactive Euler equations. The analysis aims to capture the coupling between the amplification of forward travelling pressure waves and the resulting shortening of the ignition delay discussed above.  We assume that the energy release rate is slow, i.e. 
\begin{equation}
\zeta \equiv \frac{K}{\epsilon}=O(1)
\end{equation}
This makes the solution accurate on time scales of order unity (i.e. induction time scales).  The analysis is also valid for fast heat release, which can be obtained by rescaling the time variable for faster time scales \citep{Sharpe2002}, restricting the result to earlier times.  In both cases, what we are interested in is the acceleration of the fire.  We will also restrict our analysis to the early amplification occurring before the lead shock has been modified by the arrival of a pressure wave originating in the reaction zone.  We expand the density and reaction progress variable in the form 
\begin{align}
\rho(x,t)=\rho_1(x,t)+\epsilon\rho_2(x,t)+O(\epsilon^2) \label{eq:rhoexp}\\
\lambda_r(x,t)=\lambda_{r1}(x,t)+\epsilon \lambda_{r2}(x,t)+O(\epsilon^2) \label{eq:lambdaexp}\\
t^*(x)=t_1^*(x)+O(\epsilon) \label{eq:texp}
\end{align}
Replacing \eqref{eq:rhoexp}, \eqref{eq:lambdaexp} and \eqref{eq:texp} in the governing equations and boundary conditions, we obtain to leading order:
\begin{align}
\rho_1(x,t)=1\\ 
\lambda_{r1}(x,t)=0\\
t_1^*(x)=1+2x
\end{align}\label{eq:order1}
This corresponds to the inert solution of a piston driven shock wave.

With the $O(1)$ solution obtained, the $O(\epsilon)$ problem to be solved is:
\begin{align}
\frac{d }{d t} \left( \rho_2 + \lambda_{r2} \frac{Q}{2} \right)=\frac{1}{2}\zeta Q \; \mathrm{\quad along \quad}\;  \frac{dx}{dt}=1 \label{eq:charpluseps}\\
\frac{d \lambda_{r2}}{d t}=\zeta \; \mathrm{\quad along \quad} \; \frac{dx}{dt}=0\label{eq:char0eps}\\
\rho_2(x=0,t)=0\label{eq:BCeps}\\
\lambda_{r2}(x, t<t^*)=0\label{eq:BC2eps}\\
\int_{t_s(x)}^{t*(x)} e^{\rho_2}dt=1\label{eq:inteps}
\end{align}
The resulting problem is one of linear acoustics, which is very similar in form to what is obtained for the reactive Euler case, i.e., the Clarke equations \cite{Clarkeetal1990, Sharpe2002}.  According to \eqref{eq:charpluseps}, the forward facing pressure waves all propagate at the constant speed (of unity), but amplify at constant rate in the reaction zone due to energy release.  The solution can be obtained by the method of characteristics.  In the reaction zone, for $t>t^*$ (zone R in Figure \ref{fig:fig1}), the reaction progress variable can be immediately obtained.
\begin{equation}
\lambda_{r2}(x,t>t^*)=\zeta \left(t-t^*(x)\right) \label{eq:lambdar2}
\end{equation}  
Substituting \eqref{eq:lambdar2} into \eqref{eq:charpluseps} and using the boundary condition \eqref{eq:BCeps}, we obtain the variation of density in the reaction zone:
\begin{align}
\rho_2=\frac{\zeta Q}{2}\left(t^*(x)-1\right) \label{eq:densityR}
\end{align}
Once the density is known in the reaction zone, the value of density in the disturbed region D (see Figure \ref{fig:fig1}) can be found by extending the $C^+$ characteristics into this region.   In region D, $\lambda_2=0$ and the $C^+$ characteristic relation \eqref{eq:charpluseps} requires that density is constant along a characteristic. The constant is evaluated using \eqref{eq:densityR} at the intersection of the $C^+$ characteristic with the path $t=t^*(x)$.  Denoting this reference position $x_{ref}(x,t)$ (see Figure \ref{fig:fig1}), we get:
\begin{align}
\rho_2(x, t)= \frac{\zeta Q}{2}\left(t^*(x_{ref}(x,t)-1\right) \label{eq:densityD}
\end{align}
with $x_{ref}$ given implicitly from the trajectory of the $C^+$ characteristic, i.e.,
\begin{align}
\frac{x-x_{ref}}{t-t^*(x_{ref})}=1 \label{eq:traj}
\end{align}
Region S is bounded by the shock $t_s(x) = 2x$ and the first $C^+$ disturbance originating at $(x=0,t=1)$, given by $t=1+x$.  In this region, using the method of characteristics, we get $\rho_2(x, t)= \lambda_2(x,t)= 0$.  The solution is now complete, and the integral relation \eqref{eq:inteps} can now be written as
\begin{align}
\int_{1+x}^{t^*(x)} e^{\frac{Q\zeta}{2}\left(t^*(x_{ref}(x,t)-1\right)}dt - x=0\label{eq:intfinal}
\end{align}
This expression, along with \eqref{eq:traj}, give the trajectory $t^*(x)$ implicitly.  Adopting Sharpe's iterative strategy \cite{Sharpe2002} to find $t^*(x)$, we substitute the first order approximation $t_1^*(x)$ in the integrand and solve for the correction appearing in the integral's upper bound.  From the first iteration, we find the following approximation: 
\begin{align}
t^*(x)\approx 1+x+\frac{1}{Q\zeta}\log\left( 1+Q \zeta x\right)\label{eq:tstarfinal}
\end{align}
This expression is shown in Figure \ref{fig:K=0.2}; it reproduces quite accurately the acceleration of the reaction path.  Evaluating the acceleration at the origin, we obtain:
\begin{align}
a=\frac{Q\zeta}{8} = \frac{1}{8}\frac{K}{\epsilon}Q=\frac{1}{8}\chi Q \label{eq:accel}
\end{align}
This concludes our analysis, which shows quantitatively the feedback mechanism discussed in the previous section.  More importantly, however, it shows explicitly that the acceleration of the reaction zone is given as the product of $\chi$ and the heat release Q.   
\section{Influence of $\chi$ and ignition regimes}
\begin{figure}
\begin{center}
 \includegraphics[width=110mm]{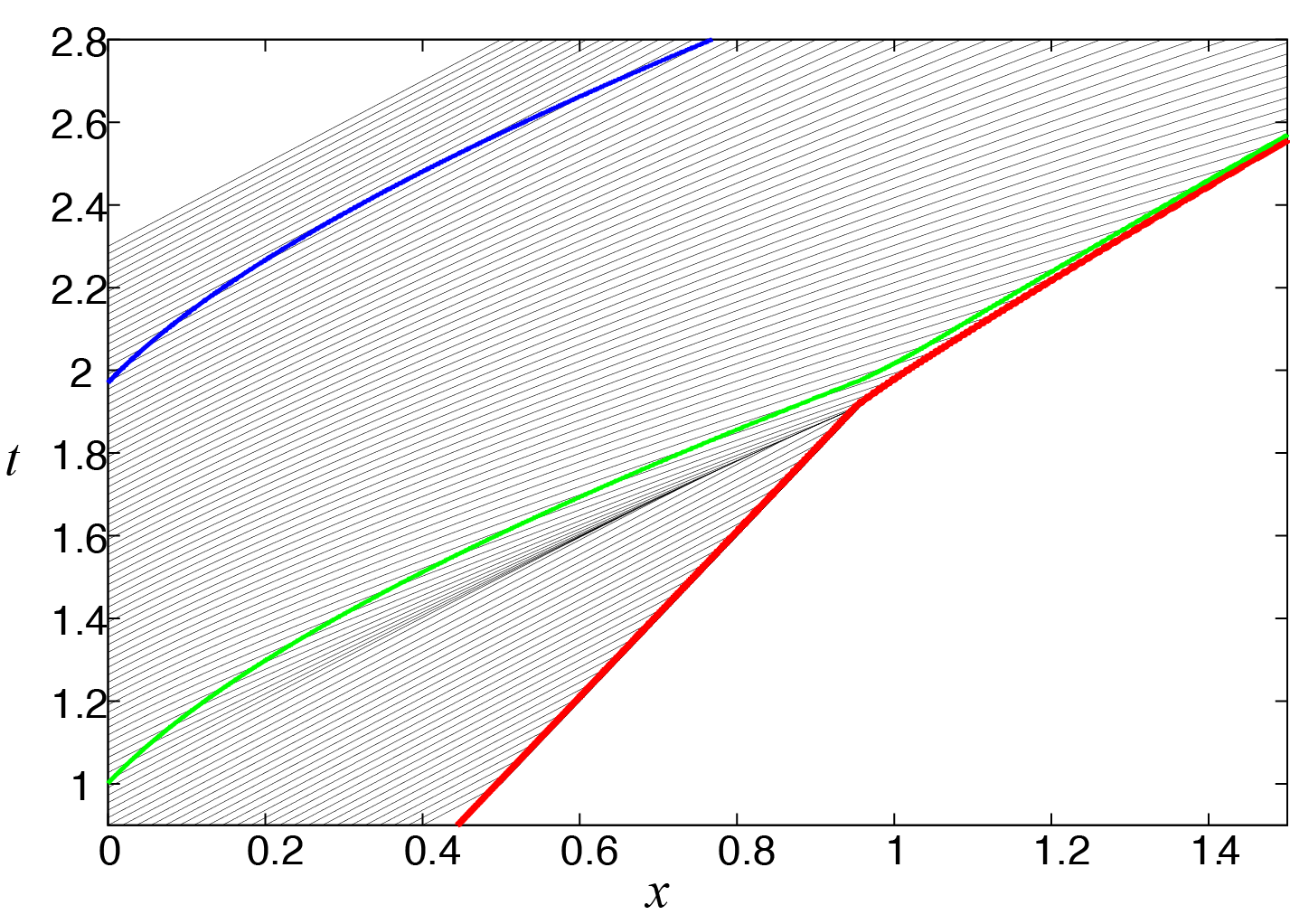}
\caption{Characteristic $t-x$ plot of shock initiation for $K$=2, $\epsilon$=0.2; same legend as with Fig. \ref{fig:K=0.2}, additionally with blue line representing the end of the reaction layer.} 
\label{fig:K2}
\end{center}
\end{figure}

\addvspace{10pt}
To illustrate the effect of varying $\chi$ on the process of ignition and detonation formation, we have conducted numerical simulations by varying both $\epsilon$ and $K$.  This permits us to determine the evolution of the flow field on longer time scales, when the analysis presented above breaks down (i.e. $\chi>>1$).  Keeping $\epsilon =0.2$ as above and increasing $K$ to a value of 2 provides a more rapid amplification, as shown in Figure \ref{fig:K2}. The accompanying \textit{Video 2} shows the evolution of the $\rho$ field.  This corresponds to $\chi=10$.  As can be seen in the figure, a shock wave now forms inside the induction zone from the confluence of the forward travelling pressure waves.  This is due to the stronger amplification of the pressure waves in the reaction zone.  This subsequent motion of the internal shock and fire are now in phase, and propagate quasi-steadily.  The model thus recovers the internal reaction wave dynamics observed experimentally \cite{Strehlow1968} and for the reactive Euler equations \citep{Sharpe2002}, where both subsonic and supersonic waves can be established.  The supersonic waves occur when the initial acceleration of the fire is sufficiently large.  

\begin{figure}
\begin{center}
 \includegraphics[width=110mm]{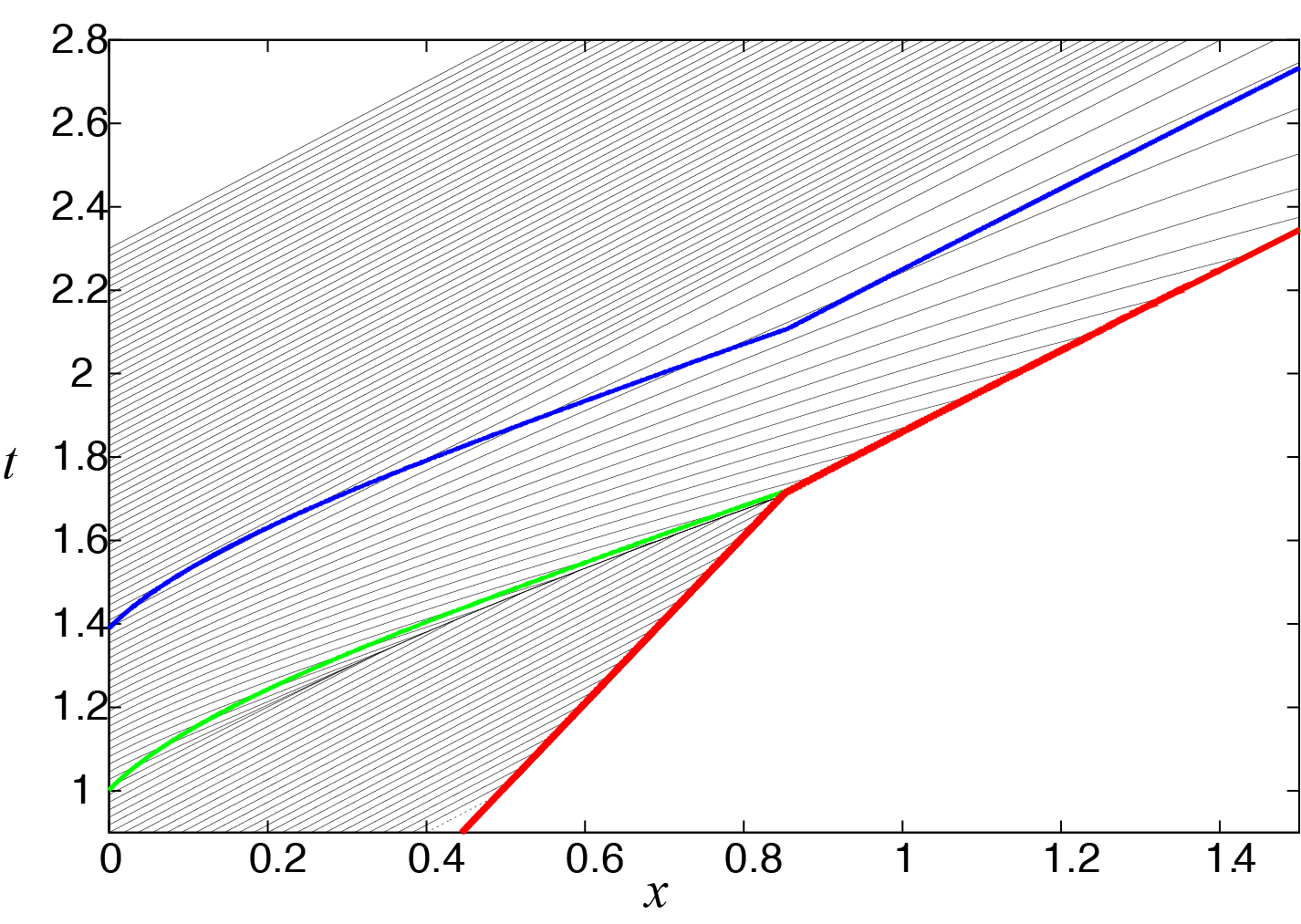}
\caption{Characteristic $t-x$ plot of shock initiation for $K$=5, $\epsilon$=0.2; same legend as for Fig. \ref{fig:K2}.} 
\label{fig:K5}
\end{center}
\end{figure}

Further increase in the rate of energy release rate $K$ leads to a more prompt amplification of the pressure waves.  Figure \ref{fig:K5} shows the evolution of the flow field when $K=5$, i.e. $\chi=50$.  The accompanying \textit{Video 3} shows the evolution of the $\rho$ field. The internal shock wave now forms at an earlier time.  The stronger compression in the induction zone now gives rise to a significant reduction in the induction delay.  The onset of energy release now becomes in phase with the internal shock motion, and the two slowly accelerate.  The arrival of this internal shock to the lead shock transforms the lead shock into a detonation wave.

This interior wave takes the form of a quasi-steady weak internal detonation.  It is not self-sustaining, as the expiration of the induction delay essentially sets its trajectory.  For reference, a Chapman-Jouguet internal wave propagating into the induction zone medium ($\rho=1$) would have a speed of 2 \cite{Fickett1985}. Instead, we find that the speed of this internal detonation wave is approximately 1.5 when it encounters the leading shock.  For even larger values of $K$, we see the same evolution, but the internal weak detonations take on velocities higher than the corresponding CJ values.

\begin{figure}
\begin{center}
 \includegraphics[width=110mm]{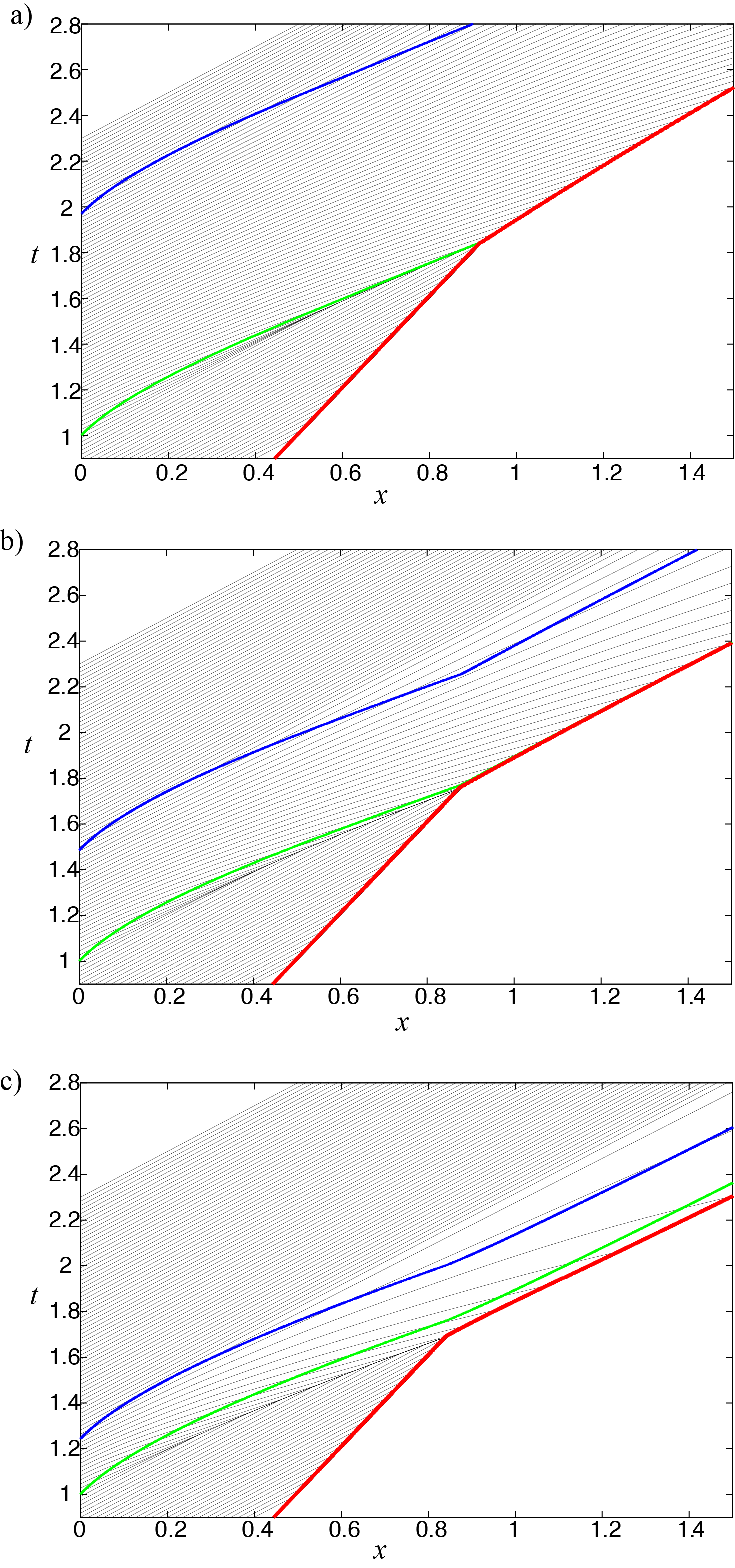}
\caption{Characteristic $t-x$ plot of shock initiation for $\chi=20$, a) $K=2$, $\epsilon$=0.1. b) $K=4$, $\epsilon$=0.2. c) $K=8$, $\epsilon$=0.4.  } 
\label{fig:epsilon}
\end{center}
\end{figure}

We have further explored the effect of varying both $\epsilon$ and $K$ in order to determine if a unique value of $\chi$ can be used to characterize the acceleration process.  While this is so for $\chi = O(1)$ through the analysis presented above, larger values of $\chi$ yield internal shock waves, invalidating the analysis once the shock has formed.  Figure \ref{fig:epsilon} shows three different cases where both $\epsilon$ and $K$ are varied so as to keep a constant $\chi=20$.  For all three cases, we see that the formation of the internal shock and subsequent amplification are very similar, in spite of large differences in reaction rates.  This suggests that indeed the entire acceleration process can be characterized by the unique parameter $\chi$.  The analysis of the acceleration of the fire in the presence of the internal shock is left for future study.    

\section{Concluding remarks}
\addvspace{10pt}
Modelling the shock induced ignition using Fickett's model permitted us to clearly illustrate the mechanism of acceleration of the induced hot spot: the energy release along the particle paths provides an amplification of forward-travelling pressure waves.  These waves pre-compress the induction layer ahead of the reaction zone, therefore changing the induction delays.  The variation of the induction delay then provides a modulation of the amplification of forward travelling pressure waves.  The model simplicity also allowed us to obtain an analytical solution for the acceleration of the reaction zone, which was found proportional to the product of the activation energy, the ratio of the induction to reaction time and the heat release.  This finding provides a theoretical justification for the previous use of this non-dimensional number to characterize the ignition regimes observed experimentally in detonations and shock induced ignition phenomena.
 
\section*{Acknowledgements}
We wish to thank the financial support of NSERC through a Discovery Grant to MIR and the support of the H2CAN NSERC Strategic Network of Excellence.


\bibliographystyle{elsarticle-num}
\bibliography{references}







\end{document}